\documentclass[aps,preprint,prd,showpacs,nofootinbib]{revtex4}

\usepackage{amsmath}
\usepackage{graphicx}
\usepackage{subfigure}
\usepackage{bm}
\usepackage{amssymb}
\usepackage{mathtools}
\usepackage{enumerate}
\usepackage{color}
\usepackage[colorlinks,linkcolor=magenta,anchorcolor=cyan,citecolor=blue]{hyperref}

\begin{document}

\title{ Quantum decoherence of primordial perturbations\\ through nonlinear scaler-tensor interaction}

\author{Gen Ye$^{1}$\footnote{yegen14@mails.ucas.ac.cn}}
\author{Yun-Song Piao$^{1,2}$\footnote{yspiao@ucas.ac.cn}}

\affiliation{$^1$ School of Physics, University of Chinese Academy of
Sciences, Beijing 100049, China}

\affiliation{$^2$ Institute of Theoretical Physics, Chinese
Academy of Sciences, P.O. Box 2735, Beijing 100190, China}

\begin{abstract}

Scaler and tensor perturbations couple nonlinearly with each other
in the Einstein-Hilbert action. We show that such interaction
naturally leads to the quantum decoherence of the primordial
perturbations during inflation at horizon crossing. The dominant
interaction Hamiltonian contributing to decoherence is identified
and the master equation responsible for the decohering process is
derived.


\end{abstract}

\maketitle


\section{ Introduction}
Inflation is a successful picture of the early universe. It solves the flatness problem, the horizon problem, and the monopole problem that shadow big bang theory\cite{Guth:1980zm,Starobinsky:1980te,Albrecht:1982wi,Linde:1981mu}. In particular, inflation states that the large scale structure of our universe is seeded by primordial quantum fluctuations. Assuming the quantum origin of our universe, a natural question to ask is:
\begin{center}
    \textit{What mechanism is responsible for the quantum to classical transition of the primordial perturbations?}
\end{center}
This is often called the measurement problem of quantum mechanics. It's usually addressed in QM textbooks by the postulate of wave function collapse (e.g: \cite{Shankar:1994principle}). Such practice turns out to be problematic in a cosmological setup\cite{Hartle:2018quantum}.

One possible solution is quantum decoherence. Quantum decoherence originated from the study of open quantum systems \cite{Zeh:1970interpretation,Joos:1985emergence}. For reviews in this field, see \cite{Zurek:2003decoherence,Schlosshauer:2005decoherence}. The idea is that through interaction with some sort of environment, the subsystem's density matrix becomes diagonal under a physically-selected basis (the pointer basis), hence classical probability is restored \cite{Zurek:1981pointer,Zurek:1982environment,Zurek:1993preferred}. In contrary to classicality, quantum probability involves interference. For example, in a double-slit experiment, it is impossible to reconstruct a classical history from the interference pattern on the screen to say which slit the electron has passed. This system decoheres if one put detectors at the slits to record electron behavior, in which case interference disappears and one may determine through which slit the electron flew. We've been reconstructing classical histories of the universe ever since the beginning of cosmology. Though supported by observation, theoretical justification is needed for this practice. Many have constructed decoherence models to address this question (e.g: \cite{Lombardo:2005decoherence,Burgess:2006primordial,Martineau:2007decoherence,prokopec2007decoherence,Koksma:2010decoherence,franco2011decoherence,Burgess2015:eft,Markkanen:2016decoherence,Nelson:2016quantum,liu2016cosmic,alexander2016inflation,Boddy:2017how,Hollowood:2017decoherence,Martin:2018observational,Martin:2018gaussianities}).

In this paper, we propose a decoherence model in a flat inflationary universe. We take the scaler perturbation of the FRW (Newtonian gauge) metric \cite{friedmann1922Zuber,lemaitre1931expansion,robertson1935kinematics,walker1937milne} as the system of interest and treat tensor perturbations as an environment. A master equation is derived governing time evolution of the density operator of the subsystem. Three points distinguish our choice of system and environment from other models:
\begin{enumerate}
    \item[(a)] Scaler-tensor coupling naturally arises from gravitational nonlinearities in GR.
    \item[(b)] Tensor perturbation is a special environment because it does not exhibit $\delta(t-t')$ form of time correlation.
    \item[(c)] The scaler system also couples with the perturbations of the inflation field.
\end{enumerate}
Similar choice of system and environment has been considered in \cite{franco2011decoherence} within the formalism of influence functions, where the tensor environment is treated in a totally stochastic way. In our paper, however, we first prove that decoherence is driven by only one of the interactions in point (c) (i.e: gravitational nonlinearity) and then derive our master equation in a non-stochastic environment (point (b)). Another feature of our work is that the final equation manifests in a functional form, which is more numerically viable than influence functions. Though there is standard derivation of master equations during inflation epoch \cite{Burgess:2006primordial,Burgess2015:eft}, such derivation does not apply to our choice of system due to point (b).

This paper is structured as follows. In section \ref{sec:scaler tensor interaction} and \ref{sec:interaction hamiltonian}, we identify the dominant interaction Hamiltonian contributing to decoherence using quantized free fields. Then we derive the master equation in section \ref{sec:derive the master equation}, with particular emphasis on the assumptions we make. In the last section \ref{sec:solution} we turn our master equation into a functional form and conclude that decoherence can occur at Hubble crossing.

\section{Interactions during inflation}\label{sec:scaler tensor interaction}
\subsection{Interaction Lagrangian}
The base metric we use is the flat FRW metric under Newtonian gauge
\begin{equation}\label{base gauge}
ds^2=a^2[-(1+2\phi)dt^2+(1-2\phi)|d\bm{x}|^2+h_{ij}dx_idx_j]
\end{equation}
$\phi$ denotes scaler metric perturbation and $h_{ij}$ is the traceless symmetric transverse tensor perturbation. The tensor modes couple with scaler metric perturbations ($\mathcal{L}_{S-T \ int}$) as well as fluctuations of the inflation field ($\mathcal{L}_{\varphi-T \ int}$). The interaction Lagrangians can be obtained by expanding the perturbed action 
\begin{equation}\label{interaction lagrangian gravity}
\mathcal{L}_{S-T \ int}=\kappa M_p^2 a^2h_{ij}\phi_{,i}\phi_{,j}
\end{equation}
\begin{equation}\label{interaction lagrangian inflation}
\mathcal{L}_{\varphi-T \ int}=\frac{1}{2}a^2h_{ij}\delta\varphi_{,i}\delta\varphi_{,j}
\end{equation}
$\kappa=1$ for GR nonlinearity (Appendix \ref{apdx:derivation of interaction lagrangian}).

\subsection{Free field quantization}\label{subsec:free field quantization}
The quantization of free perturbation fields can be found in standard textbooks (e.g: \cite{Weinberg:2008cosmology,Mukhanov:2007theory}). We list here the results we will need later on.

For free tensor perturbations, let's define $u_{ij}=ah_{ij}/\sqrt{32\pi G}=M_pah_{ij}/2$. $u_{ij}$ has mode expansion
\begin{equation}\label{mode expansion of tensor field}
\hat{u}_{ij}(t,\textbf{x})=\int\frac{dk^3}{(2\pi)^3}\sum_{s=+, \times}(\epsilon_{ij}^s(\bm{k})u^s_{k}(t)a^s_{\bm{k}}e^{i\bm{k}\cdot \bm{x}}+\epsilon_{ij}^{s*}(\bm{k})u^{s*}_k(t)a^{s\dagger}_{\bm{k}}e^{-i \bm{k}\cdot \bm{x}})
\end{equation}
with normalization relation
\begin{equation}\label{mode normalization}
2Im(u_k^su_k^{s*\prime})=\frac{1}{i}[u_k^s(u_k^{s*})'-u_k^{s*}(u_k^s)']=\frac{1}{i}\mathcal{W}(u^s_k,u^{s*}_k)=1
\end{equation}
In a slow-roll inflation background, mode function $u_k$ satisfies the equation
\begin{equation}\label{mode bessel equation}
u_k''+\left(k^2-\frac{\mu^2-1/4}{t^2}\right)u_k=0
\end{equation}
This equation has an explicit solution independent of $s$
\begin{equation}\label{tensor mode function}
u_k(t)=\frac{\sqrt{\pi}}{2}e^{i(\mu+1/2)\pi/2}\sqrt{|t|}H^{(1)}_\mu(k|t|)
\end{equation}
with $\mu=3/2+\epsilon_H$ and $\epsilon_H$ being the slow-roll parameter.

Using the gauge-invariant variable $v=a[\delta\varphi+(\varphi'/\mathcal{H})\phi]$ \cite{Mukhanov:1988vf,Kodama1984:cosmological}, the scaler field writes
\begin{equation}\label{mode expansion of scaler field}
\hat{v}(t,\textbf{x})=\int\frac{dk^3}{(2\pi)^3}(v_{k}(t)b_{\bm{k}}e^{i\bm{k}\cdot \bm{x}}+v^*_k(t)b^{\dagger}_{\bm{k}}e^{-i \bm{k}\cdot \bm{x}})
\end{equation}
\begin{equation}\label{scaler mode function}
v_k(t)=\frac{\sqrt{\pi}}{2}e^{i(\nu+1/2)\pi/2}\sqrt{|t|}H^{(1)}_\nu(k|t|)
\end{equation}
where $\nu=3/2+2\epsilon_H-\eta_H$.

Follow \cite{Burgess2015:eft} one writes the ground state wave functional (BD vacuum initial states) for the scaler field as
\begin{equation}\label{gaussian ansatz}
\psi(\xi(\bm{k}))=N(\bm{k},t)\exp\left[-\frac{1}{(2\pi)^3}\omega(\bm{k},t) \xi(\bm{k})\xi^*(\bm{k})\right]
\end{equation}
For de Sitter space-time, in particular,
\begin{equation}\label{wave functional mode dS}
\omega(k,t)=k\left(\frac{(kt)^2+i/(kt)}{(kt)^2+1}\right)
\end{equation}

\section{The interaction Hamiltonian}\label{sec:interaction hamiltonian}
We will keep track of dimension in this section for latter convenience. There are mainly two kinds of scaler perturbations coupling with the environment (i.e: tensor perturbations of the metric) during inflation:\begin{enumerate}
    \item Scaler perturbations of the metric.
    \item Fluctuations of the inflation field.
\end{enumerate}
\eqref{interaction lagrangian gravity} and \eqref{interaction lagrangian inflation} imply that one should seek a interaction term of form
\[\mathcal{L}_{int}=\xi u_{ij}v_{,i}v_{,j}\]
where $\xi$ is a coupling constant to be determined and $v=a[\delta\varphi+\phi\varphi'/\mathcal{H}]$ is the Mukhanov-Sasaki variable. Here $\mathcal{H}=a'/a$ is the comoving Hubble parameter. For gravitational interaction \eqref{interaction lagrangian gravity}, the Hamiltonian writes
\[\mathcal{H}_{int \ gra}=-\mathcal{L}_{int \ gra}=-\kappa M^2_p a^2h_{ij}\phi_{,i}\phi_{,j}=-2\kappa M_p a^{-1}\left(\frac{\mathcal{H}}{\varphi'}\right)^2u_{ij}v_{,i}v_{,j}\]
\[\xi_{gra}=2\kappa a^{-1}M_p\left(\frac{\mathcal{H}}{\varphi'}\right)^2\]
In the last equality we use the fact that $\delta\varphi$ is an identity operator in the space of scaler metric perturbations, so we have $\phi\sim [\mathcal{H}/(a\varphi')]v$ in sense of evaluating vacuum expectation $\langle \mathcal{H}_{int \ gra}\rangle$. Furthermore, the background inflation field is homogeneous (i.e: $\nabla \varphi=0$) so the factor $\mathcal{H}/\varphi'$ commutes with spacial derivatives. Similarly $\delta\varphi\sim v/a$ thus
\[\mathcal{H}_{int \ inf}=-a^{-1}M^{-1}_pu_{ij}v_{,i}v_{,j}\]
\[\xi_{inf}=a^{-1}M^{-1}_p\]
By comparing the coupling constants
\[\frac{\xi_{inf}}{\xi_{gra}}\approx\epsilon_H/\kappa\ll1 \]
we conclude that the dominant interaction is that between the tensor and scaler components of metric perturbations. Therefore the interaction Hamiltonian to be studied in this paper is
\[
\mathcal{H}_{int}=-2\kappa a^{-1}M_p\left(\frac{\mathcal{H}}{\varphi'}\right)^2u_{ij}v_{,i}v_{,j}=-\frac{\kappa}{a\epsilon_HM_p}u_{ij}v_{,i}v_{,j}
\]
The above argument is inspired by \cite{Burgess:2006primordial,Martineau:2007decoherence}. Note that because field operators are Heisenberg, this interaction Hamiltonian is also Heisenberg. In the following sections we will need its decomposition to the tensor and scaler part
\begin{equation}\label{interaction hamiltonian}
\tilde{H}_{int}(t)=\sum_{\bm{x}}T_{\bm{x}}(t)\otimes S_{\bm{x}}(t)
\end{equation}
and
\begin{equation}\label{T_x,S_x}
\begin{aligned}
&\tilde{T}_{\bm{x}}(t)=\int\frac{dk^3}{(2\pi)^3}\sum_{s=+,\times}\epsilon^s_{ij}(\bm{k})(u_k(t)a_{\bm{k}}^s+u_k^*(t)a_{\bm{-k}}^{s\dagger})e^{i\bm{k}\cdot\bm{x}}\\
&\tilde{S}^{(ij)}_{\bm{x}}(t)=\lambda\int\frac{dk'^3}{(2\pi)^3}\frac{dk''^3}{(2\pi)^3}(-k_i'k_j'')(v_{k'}b_{\bm{k}'}+v^*_{k'}b^\dagger_{-\bm{k}'})(v_{k''}b_{\bm{k}''}+v^*_{k''}b^\dagger_{-\bm{k}''})e^{i(\bm{k}'+\bm{k}'')\cdot\bm{x}}
\end{aligned}
\end{equation}
where $\lambda=-\kappa/(a\epsilon_HM_p)$ is of dimension $(mass)^{-1}$.

\section{Derive the master equation}\label{sec:derive the master equation}
The total Hamiltonian of the system and the environment is
\[H=H_{T}\otimes I_S+I_T\otimes H_{S}+\gamma V\]
The interaction term is contraction of three perturbation fields, thus we shall carry out our derivation in the weak coupling limit. For convenience of expansion we write out the weak coupling constant $\gamma$ explicitly.

We start with the Liouville-von Neumann equation in the interaction picture\footnote{Interaction picture actually requires a time-independent free Hamiltonian. It's not the case here. However, this is no problem because all we need is the operator equation \[i\frac{\partial U}{\partial t}(t_0;t)=H(t)U(t_0;t)\] which holds even if $H_0$ depends on time.}
\begin{equation}\label{liouville eq}
i\frac{d\tilde{\rho}}{dt}=\gamma[\tilde{V},\tilde{\rho}]
\end{equation}
Quantities in the interaction picture are labeled by tilde
\[\tilde{\rho}(t)=U(t_0;t)^\dagger\rho(t)U(t_0;t)\qquad \tilde{V}=U(t_0;t)^\dagger V U(t_0;t)\]
\[U(t_0;t)=T\exp\left(-i\int_{t_0}^{t}H_0(t')dt'\right)=U_T(t_0;t)\otimes U_S(t_0;t)\]
Equation \eqref{liouville eq} has a formal solution
\[\tilde{\rho}(t)-\tilde{\rho}(t_0)=-i\gamma\int_{t_0}^tdt'[\tilde{V}(t'),\tilde{\rho}(t')]\]
The solution can be approximated by a Dyson series, keeping only the lowest two powers of $\gamma$
\begin{equation}\label{2 order exp}
\begin{aligned}
\frac{d\tilde{\rho}}{dt}=&-i\gamma[\tilde{V}(t),\tilde{\rho}(t_0)]-\gamma^2\int_{t_0}^{t}dt'[\tilde{V}(t),[\tilde{V}(t'),\tilde{\rho}(t')]]
\end{aligned}
\end{equation}

Let's assume a separable initial state $\tilde{\rho}(t_0)=\rho(t_0)=\varrho_T\otimes\varrho_S$ and recall the decomposed interaction term \eqref{interaction hamiltonian} then
\[
\begin{aligned}
Tr_T\left([\tilde{V}(t),\tilde{\rho}(t_0)]\right)&=\lambda Tr_T\left(\sum_i[\tilde{T}_i(t)\otimes \tilde{S}_i(t),\varrho_T\otimes\varrho_S]\right)\\&=\lambda\sum_iTr_T(\varrho_T\tilde{T}_i(t))[\tilde{S}_i(t),\varrho_S]
\end{aligned}
\]
Under particle number basis the trace is expanded by $Tr_T(\mathcal{O})=\int \mathcal{D}[n(\bm{k})]\langle\bm{n}| \mathcal{O}|\bm{n}\rangle$
Note that $\langle n|\cdot|n\rangle$ and $\varrho_T$ both provide even number of creation/annihilation operators while $\tilde{T}_i(t)$ only contributes one, so $Tr_T(\varrho_T\tilde{T}_i(t))=0$ by normal ordering. Hence the term linear in $\gamma$ vanishes
\begin{equation}\label{master eq1}
\frac{d\tilde{\rho_S}}{dt}=-\gamma^2\int_{t_0}^{t}dt'Tr_T\left([\tilde{V}(t),[\tilde{V}(t'),\tilde{\rho}(t')]]\right)
\end{equation}
This equation means that the correction induced by interaction is second order in $\gamma$
\[\rho(t)=U_T\varrho_TU^\dagger_T\otimes U_S\varrho_SU^\dagger_S-\gamma^2\rho_c(t)\]
In the interaction picture
\[\tilde{\rho}(t)=\varrho_T\otimes\varrho_S-\gamma^2\tilde{\rho}_{c}(t)\]
The correction $\tilde{\rho_c}$ carries $\gamma^4$ in the RHS of equation \eqref{master eq1}, so one can literally use any time $t_0<t'<t$ as the time argument of $\tilde{\rho}$. Since we want to keep track of the scaler part, let's make the substitution $\tilde{\rho}(t')\to\tilde{\rho}_S(t)\otimes\varrho_T$ in equation \eqref{master eq1} and expand the trace term
\[
\begin{aligned}
Tr_T&([V(t),[V(t'),\rho(t')]])=\sum_{i,j}Tr_T\left[T_i\otimes S_i,\left[T_j\otimes S_j,\rho_S\otimes\varrho_T\right]\right]\\&=\sum_{i,j}\left[Tr_T(T_iT_j\varrho_T)(S_iS_j\rho_S-S_j\rho_SS_i)+Tr_T(T_jT_i\varrho_T)(\rho_SS_jS_i-S_i\rho_SS_j)\right]\\&=\sum_{\bm{x},\bm{x}'}Tr_T(T_{\bm{x}}T_{\bm{x}'}\varrho_T)(S_{\bm{x}}S_{\bm{x}'}\rho_S-S_{\bm{x}'}\rho_SS_{\bm{x}})+h.c.
\end{aligned}
\]
where time arguments $t,t'$, upper indexes $(ij)$ of $S$ and tilde for interaction picture are omitted. $h.c.$ stands for Hermitian Conjugate. Now it remains to calculate partial trace $Tr_T(T_{\bm{x}}T_{\bm{x}'}\varrho_T)$. Under certain assumptions, the partial trace can be approximated by a tensor acting as some sort of propagator on the horizon (see Appendix \ref{apdx:calculate trace tensor})
\[ Tr_T(T_{\bm{x}}(t)T_{\bm{x}'}(t')\varrho_T)\equiv T_{ijkl}\delta(t-t'-\Delta x)/\Delta x\]
where
\begin{equation}\label{T trace tensor}
T_{ijkl}=\begin{bmatrix}
\begin{smallmatrix}T_1&0&0\\0&T_2&0\\0&0&T_5\end{smallmatrix}
&\begin{smallmatrix}0&T_3&0\\T_3&0&0\\0&0&0\end{smallmatrix}
&\begin{smallmatrix}0&0&T_4\\0&0&0\\T_4&0&0\end{smallmatrix}\\
\begin{smallmatrix}0&T_3&0\\T_3&0&0\\0&0&0\end{smallmatrix}
&\begin{smallmatrix}T_2&0&0\\0&T_1&0\\0&0&T_5\end{smallmatrix}
&\begin{smallmatrix}0&0&0\\0&0&T_4\\0&T_4&0\end{smallmatrix}\\
\begin{smallmatrix}0&0&T_4\\0&0&0\\T_4&0&0\end{smallmatrix}
&\begin{smallmatrix}0&0&0\\0&0&T_4\\0&T_4&0\end{smallmatrix}
&\begin{smallmatrix}T_5&0&0\\0&T_5&0\\0&0&-2T_5\end{smallmatrix}
\end{bmatrix}
\end{equation}
is dimensionless. $\Delta x = |\bm{x}-\bm{x}'|$ and
\[\begin{aligned}
&T_1=\frac{1}{8\pi^2}\left(\frac{3}{4}\alpha_1+p-i\frac{\pi}{2}\right) \qquad T_2=\frac{1}{8\pi^2}\left(-\frac{13}{12}\alpha_2-p+i\frac{\pi}{2}\right) \qquad T_4=\frac{1}{8\pi^2}\left(-\frac{2}{3}\alpha_4\right)\\
&T_3=(T_1-T_2)/2\qquad T_5=-T_1-T_2
\end{aligned}\]
We list  some important properties of the partial trace here
\begin{enumerate}
    \item Nonzero entries of $T_{ijkl}$ are proportional to $(\delta_{ik}\delta_{jl}+\delta_{il}\delta_{jk})$ or $\delta_{ij}\delta_{kl}$;
    \item The dominant interaction propagates at the speed of light;
    \item Expression \eqref{T trace tensor} is written in a Cartesian frame with $\Delta \bm{x}$ being its z-axis (i.e: $T_{ijkl}=T_{ijkl}(\Delta \bm{x})$).
\end{enumerate}
$\delta$ functions cancel out the time integration and we get our final result in this section
\begin{equation}\label{The Master Eq Heinsenberg}
\begin{aligned}
\frac{d\tilde{\rho}_S}{dt}=-\sum_{\bm{x},\bm{x}'}\frac{1}{\Delta x}T_{ijkl}\left(S^{(ij)}_{\bm{x}}(t)S^{(kl)}_{\bm{x'}}(t-\Delta x)\tilde{\rho}_S(t)-S^{(kl)}_{\bm{x'}}(t-\Delta x)\tilde{\rho}_S(t)S^{(ij)}_{\bm{x}}(t)\right)+h.c.
\end{aligned}
\end{equation}
where $\hat{S}^{(ij)}_{\bm{x}}=\lambda \hat{v}_{,i}\hat{v}_{,j}$ can be expanded in momentum space \eqref{T_x,S_x}.

\section{Solution}\label{sec:solution}
\subsection{General discussions}
Interaction \eqref{interaction hamiltonian} is diagonal in configuration space. This implies that the natural pointer basis for the system is the field amplitude basis $\{|\bm{v}\rangle\}$
\[\hat{v}(\bm{x})|\bm{v}\rangle=v(\bm{x})|\bm{v}\rangle\]
See Appendix \ref{apdx:field amplitude basis} for a detailed construction of this basis. Such implication coincides with our expectation, that the quantum fluctuations evolve to classical perturbation fields.

We will restrict ourselves to real fields afterwards. Under the real field amplitude basis, we can write equations of matrix elements of the system's density operator
\begin{equation}\label{The Master Eq matrix}
\begin{aligned}
\langle \bm{\xi}(t)|\frac{\partial}{\partial t}\tilde{\rho}_S(t)|\bm{\zeta}(t)\rangle=-\lambda\sum_{\bm{x},\bm{x}'}&\frac{1}{\Delta x}\left(\partial_i\xi(\bm{x})\partial_j\xi(\bm{x})-\partial_i\zeta(\bm{x})\partial_j\zeta(\bm{x})\right)\\&\times\left(T_{ijkl}\langle\bm{\xi}|S^{(kl)}_{\bm{x}'}(t-\Delta x)\tilde{\rho}_S|\bm{\zeta}\rangle-T^*_{ijkl}\langle\bm{\xi}|\tilde{\rho}_SS^{(kl)}_{\bm{x}'}(t-\Delta x)|\bm{\zeta}\rangle\right)
\end{aligned}
\end{equation}
This equation is still quite hard to use in the practical sense. Our next task is to linearize and simplify it to a functional equation of classical field configurations $\xi(\bm{x})$ and $\zeta(\bm{x})$.
\subsection{Linearized master equation}\label{sec:solution:subsec:linearization}
\paragraph*{Assumption: Linearized time evolution}\textit{We introduce the interaction window $|\Delta k/k_*|,|\Delta x/t|<\Delta\ll1$. }
$k_*$ is the mode of interest. The physics behind the constraint on $k$ is that a mode cannot feel curvature perturbations much smaller or larger than its own wavelength. Influence from distant past may be screened by other QFT interactions when traveling through space-time, so we also introduce a time window. Relaxing the time window constraint will result in more rapid decherence. Expand mode functions \eqref{scaler mode function} near conformal time $t$
\begin{equation}\label{linearized time windowed int}
v_k(t-\Delta x)\approx v_k(t)-v'_k(t)\Delta x= v_k(t)+(-|t|v'_k(t))\frac{\Delta x}{|t|}
\end{equation}
Inserting into expansion \eqref{T_x,S_x}, one obtain in the linear regime
\[\begin{aligned}
S^{(kl)}_{\bm{x}'}(t-\Delta x)=S^{(kl)}_{\bm{x}'}(t)&+\lambda\frac{\Delta x}{|t|}\int\frac{dk^3}{(2\pi)^3}(ik_k)(-|t|v'_k(t),-|t|v'^*_k(t))\frac{1}{i}\begin{bmatrix}
v^{\prime*}_k&-v^*_k\\-v'_k&v_k
\end{bmatrix}\begin{bmatrix}
\mathcal{F}[\hat{\bm{v}}](\bm{k})\\\mathcal{F}[\hat{\bm{\pi}}](\bm{k})
\end{bmatrix}e^{i\bm{k}\cdot\bm{x}'}\hat{v}_{,l}(\bm{x}')\\&+\lambda\frac{\Delta x}{|t|}\hat{v}_{,k}(\bm{x}')\int\frac{dk^3}{(2\pi)^3}(ik_l)(-|t|v'_k(t),-|t|v'^*_k(t))\frac{1}{i}\begin{bmatrix}
v^{\prime*}_k&-v^*_k\\-v'_k&v_k
\end{bmatrix}\begin{bmatrix}
\mathcal{F}[\hat{\bm{v}}](\bm{k})\\\mathcal{F}[\hat{\bm{\pi}}](\bm{k})
\end{bmatrix}e^{i\bm{k}\cdot\bm{x}'}
\end{aligned}\]
All the quantities on the RHS are evaluated at conformal time $t$. See Appendix \ref{apdx:field amplitude basis} for definition of $\mathcal{F}[\hat{\bm{v}}]$. Define $\beta_{\nu,k}$ by
\[(-|t|v'_k(t),-|t|v'^*_k(t))\begin{bmatrix}
v^{\prime*}_k&-v^*_k\\-v'_k&v_k
\end{bmatrix}=-(0,2i|t|\Im(v_kv'^*_k))\equiv(0,-i\beta_{\nu,k})\]
Notice that $\big[\hat{v}_{,k}(\bm{x}'),\mathcal{F}[\hat{\bm{\pi}}](\bm{k})\big]=k_ke^{-i\bm{k}\cdot\bm{x}'}$ then
\begin{equation}\label{linearized S}
\begin{aligned}
S^{(kl)}_{\bm{x}'}(t-\Delta x)&=S^{(kl)}_{\bm{x}'}(t)-i\lambda\frac{\Delta x}{|t|}\left[2\int\frac{dk^3}{(2\pi)^3}\beta_{\nu,k}\mathcal{F}[\hat{\bm{\pi}}](\bm{k})e^{i\bm{k}\cdot\bm{x}'}k_k\hat{v}_{,l}(\bm{x}')+\int\frac{dk^3}{(2\pi)^3}\beta_{\nu,k}k_kk_l\right]\\
&=S^{(kl)}_{\bm{x}'}(t)-i\lambda\frac{\Delta x}{|t|}\left[2\hat{v}_{,k}(\bm{x}')\int\frac{dk^3}{(2\pi)^3}\beta_{\nu,k}\mathcal{F}[\hat{\bm{\pi}}](\bm{k})e^{i\bm{k}\cdot\bm{x}'}k_l-\int\frac{dk^3}{(2\pi)^3}\beta_{\nu,k}k_kk_l\right]
\end{aligned}
\end{equation}
where we used the fact that indexes $(k,l)$ are symmetric. The term coming from commutation satisfies
\[\int\frac{dk^3}{(2\pi)^3}\beta_{\nu,k}k_kk_l\propto\delta_{kl}\]
So it vanishes since $T_{ijkl}$ is a traceless tensor. Let's adopt the assumption
\begin{center}
    \textit{The perturbation field $v(\bm{x})$ is largely homogeneous and isotropic due to inflation.}
\end{center}
So that $\mathcal{F}[\hat{\bm{\pi}}](\bm{k})\to\mathcal{F}[\hat{\bm{\pi}}](k)$ commutes with the angular integral
\[\frac{1}{(2\pi)^3}\int_{\Delta k/k_*<\Delta}dkk^3\beta_{\nu,k}\mathcal{F}[\hat{\bm{\pi}}](k)
\int d\hat{\bm{k}}\hat{\bm{k}}_le^{ik_*\hat{\bm{k}}\cdot \bm{x}'}\]
The angular part yields
\begin{equation}\label{W definition}
\frac{4\pi i}{k_*x'}\left(\frac{\sin(k_*x')}{k_*x'}-\cos(k_*x')\right)\hat{\bm{x}}'_l\equiv 4\pi i W(k_*x')\hat{\bm{x}}_l'
\end{equation}

Collect all the results and change to integration variables $\bm{r}'=\Delta \bm{x}=\bm{x}-\bm{x}'$ and $\bm{r}=\bm{x}'$, equation \eqref{The Master Eq matrix} then reduces to
\begin{equation}\label{The Master Eq matrix simplified}
\begin{aligned}
\langle\bm{\xi}|\frac{\partial}{\partial t}\rho_S|\bm{\zeta}\rangle=-\lambda^2\int d&\bm{r}d\bm{r}'\frac{1}{r'}\partial^{(-)}_{ij}[\xi,\zeta](\bm{r}+\bm{r}')\Bigg[\Re(T_{ijkl})\partial^{(-)}_{kl}[\xi,\zeta](\bm{r})+i\Im(T_{ijkl})\partial^{(+)}_{kl}[\xi,\zeta](\bm{r})\\&-i\frac{r'}{|t|}W(k_*r)\hat{\bm{r}}_k\big(T_{ijkl}\partial_{l}\xi(\bm{r})D_{\xi}+T^*_{ijkl}\partial_{l}\zeta(\bm{r})D_\zeta\big)\Bigg]\langle\bm{\xi}|\rho_S|\bm{\zeta}\rangle
\end{aligned}
\end{equation}
where
\[\partial^{(\pm)}_{ij}[\xi,\zeta](\bm{x})=\partial_i\xi(\bm{x})\partial_j\xi(\bm{x})\pm\partial_i\zeta(\bm{x})\partial_j\zeta(\bm{x})\]
\[D_{\xi}=\int_{\Delta k/k<\Delta}\frac{dk}{\pi^2}k^3\beta_{\nu,k} \mathcal{F}[\frac{d}{d\xi}](k)
\]
\subsection{Decoherence}
We first make a convenience choice of initial state. Let $\tau$ be the proper time then conformal time $t\equiv -\int^0_{\tau}d\tau'/a(\tau)$ and the initial $\tau =0$ state is set to be the end of inflation, thus $t=(a_0H)^{-1}-(aH)^{-1}= (aH)^{-1}(a/a_0-1)$. Since $a/a_0$ is exponentially small during inflation, one has $t\simeq-(aH)^{-1}$ \cite{Riotto:2002inflation}\footnote{This choice of initial state coincides with the explicit mode functions \eqref{tensor mode function}.}.

To qualitatively estimate the decoherence rate, it's most convenient to work with dimensionless variables
\[\bm{K}=\bm{k}/k_*\qquad \bm{s}=k_*\bm{r}\qquad k_*=a(t_*)H\]
and normalized fields \cite{Ewan:1993analytic}
\[\xi=\frac{aH}{2\pi}\bar{\xi} \qquad \zeta=\frac{aH}{2\pi}\bar{\zeta}\]
By "normalized" we mean bn fields $\bar{\xi},\bar{\zeta}$ are of unity amplitude. With these we can rewrite \eqref{The Master Eq matrix simplified}
\begin{equation}\label{The Master Eq dimensionless}
\begin{aligned}
\langle\bar{\xi}|\rho'_S|\bar{\zeta}\rangle=-\frac{\kappa^2}{(2\pi)^4\epsilon^2_H}\left(\frac{H}{M_p}\right)^2\left(\frac{t_*}{-t^2}\right)\int_{L_0} d\bm{s}&\int_{\Delta}d\bm{s}'\bar{\partial}^{(-)}_{ij}[\bar{\xi},\bar{\zeta}]\Bigg[\frac{1}{s'}\left(\Re(T_{ijkl})\bar{\partial}^{(-)}_{kl}[\bar{\xi},\bar{\zeta}]+i\Im(T_{ijkl})\bar{\partial}^{(+)}_{kl}[\bar{\xi},\bar{\zeta}]\right)\\&- iW(s)\hat{\bm{s}}_l(T_{ijkl}\bar{\partial}_k\bar{\xi}\bar{D}_{\bar{\xi}}+T^*_{ijkl}\bar{\partial}_k\bar{\zeta}\bar{D}_{\bar{\zeta}})\Bigg]\langle\bar{\xi}|\rho_S|\bar{\zeta}\rangle
\end{aligned}
\end{equation}
where $T_{ijkl}=T_{ijkl}(\bm{s}')$ while all the other quantities on the RHS are evaluated at $\bm{s}$. In particular, we changed $\partial^{(-)}_{ij}[\xi,\zeta](\bm{r}+\bm{r}')$ in \eqref{The Master Eq matrix simplified} to $\bar{\partial}^{(-)}_{ij}[\bar{\xi},\bar{\zeta}](\bm{s})$ here because the $\bm{s}'$ variation does not contribute in linear order (Appendix \ref{apdx:integration involving T}). The spacial cutoff $L_0$ is also a time cutoff. Due to inflation, the mode of interest evolves all the way from sub-Hubble (i.e: $k_*\gg a(t_i)H$) to its Hubble crossing (i.e: $k_*=a(t_*)H=\mathcal{H}_*$), and $L_0(t)$ can be seen as the particle horizon at time $t$, starting from $t_i$. Then $L_0(t)=k_*(t-t_i)\simeq t_i/t_*$.  The dimensionless normalized $\bar{D}_{\bar{\xi}}$ is defined as
\[\bar{D}_{\bar{\xi}}=\int \frac{dK}{\pi^2}K^3\bar{\beta}_{\nu,K}\bar{\mathcal{F}}[\frac{d}{d\bar{\xi}}](K) \]
where
\[\bar{\beta}_{\nu,K}=\frac{\pi}{4}\Im\left[\left(H^{(1)}_{\nu-1}(Ka_*/a)-H^{(1)}_{\nu+1}(Ka_*/a)\right)H^{(2)}_\nu(Ka_*/a)\right]\]
$\bar{\beta}_{\nu,K}\to0$ for $a_*/a\ll1$, hence linear correction becomes negligible when modes are sub-Hubble.

To the leading order,  \eqref{The Master Eq dimensionless} has formal solution
\[\langle\bm{\xi}|\rho_S|\bm{\zeta}\rangle\propto e^{-\int dt'\Gamma}\]
\[\Gamma\equiv\frac{\kappa^2}{(2\pi)^4\epsilon^2_H}\left(\frac{H}{M_p}\right)^2\left(\frac{t_*}{-t^2}\right)\int_{L_0} d\bm{s}\int_{\Delta}d\bm{s}'\bar{\partial}^{(-)}_{ij}[\bar{\xi},\bar{\zeta}]\frac{1}{s'}\left(\Re(T_{ijkl})\bar{\partial}^{(-)}_{kl}[\bar{\xi},\bar{\zeta}]+i\Im(T_{ijkl})\bar{\partial}^{(+)}_{kl}[\bar{\xi},\bar{\zeta}]\right)\]
The dominant time dependence in $\Gamma$ is $t^{-2}$ as other $t$ dependence are exponentially suppressed by inflation, then the time varying part of $\langle\bm{\xi}|\rho_S|\bm{\zeta}\rangle$ is approximately (for modes of scale $k_*$, $\partial_i\sim k_*=a_*H$)
\begin{equation}\label{decoherence exponential}
\exp\left[-\frac{\kappa^2\Delta^2L_0^3}{(2\pi)^4\epsilon^2_H}\left(\frac{H}{M_p}\right)^2\left(\frac{t_*}{t}\right)\Bigg|^{t_*}_{t_i}(const.)\right]\simeq\exp\left[-\frac{\kappa^2\Delta^2}{2\pi^2\epsilon_H}A_{\mathcal{R}}e^{3\Delta N}\times(const.)\right]
\end{equation}
First of all, $\bar{\partial}^{(-)}_{ij}[\bar{\xi},\bar{\zeta}]=0$ if $\xi=\zeta$, so diagonal terms of density matrix does not decay with time. Here $A_{\mathcal{R}}\sim10^{-9}$ is the amplitude of comoving curvature perturbation. $\Delta N=\log(t_i/t_*)$ is the e-folds a mode has experienced from some initial time $t_i$ till its Hubble crossing $t_*$. If $\epsilon\sim10^{-2}$ then the mode needs a few e-folds ($\Delta N\gtrsim6$) before Hubble crossing to decohere. However, we actually obtain an upper bound of decoherence time here. For example, if we relax the space-time window constraint $\Delta t/t<\Delta$, then $s'$ is also bounded by particle horizon $L_0=t_i/t_*$, hence the e-folds dependence in \eqref{decoherence exponential} becomes $e^{6\Delta N}$ and decoherence requires fewer e-folds ($\Delta N\sim3$). Actually, it might be more realistic to discard the $\Delta t/t<\Delta$ constraint in that $N\sim3$ corresponds to $0.1H\lesssim\lambda\lesssim H$, in which case gravity is indeed the dominant interaction. The positive definiteness of the real part of \eqref{decoherence exponential} can be shown by detailed computation of the $\bm{s}'$ integral (Appendix \ref{apdx:integration involving T}).

Detailed numeric simulation requires a physical coarse-graining of possible histories (i.e: sampling of the functional space of $\xi$ and $\zeta$ fields). We use a toy model to produce some visual illustration. Let's generate fields $\bar{\xi}$ and $\bar{\zeta}$ from Fourier distribution parametrized by $0<\lambda<1$
\[\xi_\lambda(K)=\max\left(0,\frac{20}{3}-\frac{400}{9}|x-1-(1-2\lambda)\Delta|\right)\]
The distribution is a peak at the $\lambda$ partition point of section $[1-\Delta,1+\Delta]$.
Keep only the leading real terms in \eqref{The Master Eq dimensionless} one then obtain the exponential suppression of non-diagonal elements (assuming a constant $p$)
\begin{equation}
\begin{aligned}
\exp\left[-\sigma e^{3\Delta N}\Big\langle integral\Big\rangle\right]
\end{aligned}
\end{equation}
where
\[L^3_0\Big\langle integral\Big\rangle=\int_{L_0} d\bm{s}F_0[\bar{\xi},\bar{\zeta}]\]
\[F_0[\bar{\xi},\bar{\zeta}]=(|\nabla\bar{\xi}|^2-|\nabla\bar{\zeta}|^2)^2+3|\nabla\bar{\xi}\times\nabla\bar{\zeta}|^2\]
\[\sigma=\frac{2p\Delta^2}{15\pi}\frac{\kappa^2}{2\pi^2\epsilon_H}A_{\mathcal{R}}\]

\begin{figure}
    \subfigure[intitial state]{\includegraphics[width=2in]{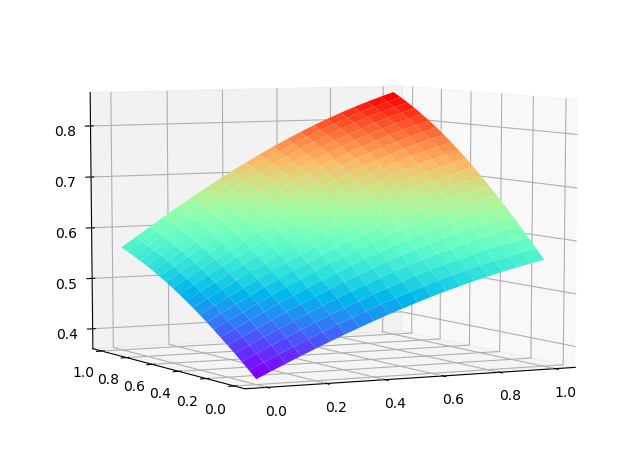}}
    \subfigure[$\sigma e^{3\Delta N}=2.4$]{\includegraphics[width=2in]{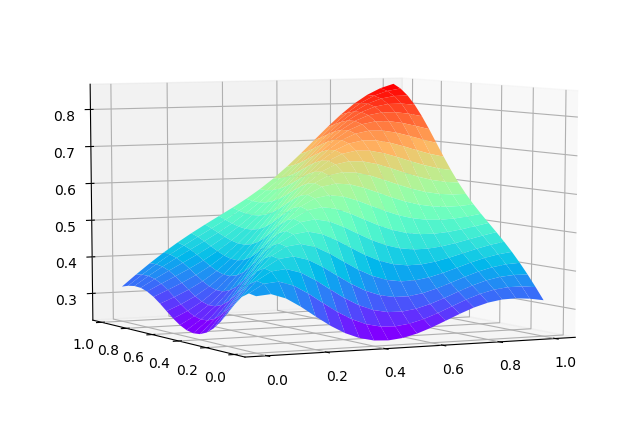}}
    \subfigure[$\sigma e^{3\Delta N}=4.8$]{\includegraphics[width=2in]{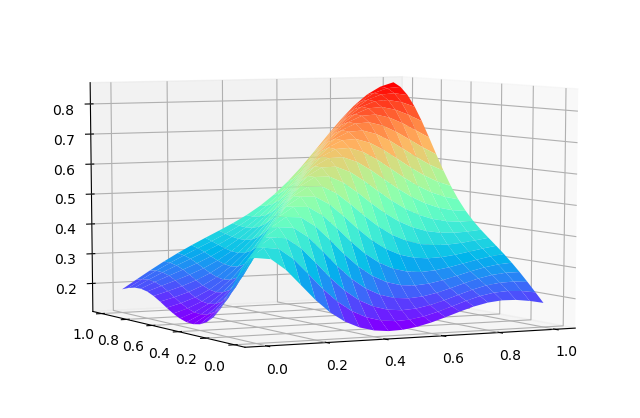}}
    \caption{Numeric illustration with a interaction window $\Delta=0.3$, $p=0.5$, x-y axis being parameter $\lambda$ for $\xi$ and $\zeta$ respectively. The scale of the $const.$ in \eqref{decoherence exponential} for off-diagonal matrix elements is $\sim 0.2$, which implies an upper bound of decoherence time in terms of e-folds $\Delta N\lesssim \frac{1}{3}\log(5/\sigma)$.}
\end{figure}

\section{Conclusion}
In this paper we argued that the decoherence of the scaler
perturbation is driven by its nonlinear coupling with the tensor
environment, rather than the quantum fluctuations of the inflation
field. We have shown that decoherence induced by gravitational
nonlinearity finishes at horizon crossing if it starts a few
e-folds earlier. Decoherence becomes more efficient if one relaxes
the linearization conditions or some interaction stronger than
gravity is present. Therefore, we reach the conclusion that
decoherence of the scaler metric perturbations completes before
modes become super-Hubble.

Note that decoherence in our model happens before the well-known "decoherence without decoherence" \cite{grishchuk1990squeezed,albrecht1994inflation,polarski1996semiclassicality,kiefer:1998quantum,jonathan:2007decoherence} mechanism takes effect, and can be tracked by a numerically viable functional equation. Our work may be a complement to the current picture of quantum to classical transition during inflation.

Despite decoherence, our work didn't find clues for possible relic of quantum origin of the universe. Also, there are technical assumptions in our derivation that may be improved (for example, $\xi$ and $\zeta$ can be complex scaler fields in general, constraint on time window can be relaxed, etc.). The master equation itself has numeric potential to be exploited. We hope our work can inspire further study in related fields.

\textbf{Acknowledgments} GY would like to thank Qing-Guo Huang and
Yu Tian for useful comments on his bachelor's thesis, and
hospitality of Columbia University in the city of New York as a
visiting student from University of Chinese Academy of Sciences.
This work is supported by NSFC, No.11575188, 11690021, and also
supported by the Strategic Priority Research Program of CAS,
No.XDB23010100.

\appendix
\section{Derivation of interaction Lagrangian}\label{apdx:derivation of interaction lagrangian}
\subsection{Interaction with scaler metric perturbation}
Because the scaler-vector-tensor decomposition of FRW metric is done on constant time hypersurfaces and a master equation has to be derived under the Hamiltonian formalism, we need a parametrization of GR that splits time and space explicitly. The ADM formalism is what we need\cite{ADM:2004dynamics}. The GR lagrangian in (3+1) dimensional space-time
\begin{equation}\label{GR larangian}
\mathcal{L}=\frac{1}{16\pi G}N\sqrt{g}\left[\prescript{(3)}{}{R}+K^{ij}K_{ij}+K^2-2(\partial_tK-\nabla^i\nabla_i N)/N\right]
\end{equation}
$K_{ij}$ is the extrinsic curvature of the constant time hypersurface. The free lagrangian of scaler and tensor perturbations are second order terms in the perturbed GR lagrangian, so the interactions between them are at least third order. The highest derivatives in \eqref{GR larangian} are second order in space, so all possible lowest order scaler-tensor interaction terms are
\[h_{ij}\phi_{|ij}\phi\quad h_{ij}\phi_{,i}\phi_{,j}\quad h_{ij}h_{ij}\phi\quad h_{ik}h_{kj}\phi_{|ij}\]
where the lower index $|$ stands for covariant derivatives correpond to the spacial metric $g_{ij}$. We can drop terms with only one scaler field because of possible complex field generalization of our theory, though we will deal with real fields only in this paper. Notice that the 3-connection $\Gamma^k_{ij}$ is at least first order in perturbation, so in the lowest order spacial covariant derivatives reduce to ordinary partial derivatives( i.e: $|\to,$). Thus we can restrict ourselves to consider only $h_{ij}\phi_{,ij}\phi$ and $h_{ij}\phi_{,i}\phi_{,j}$. These two are equivalent up to a boundary term and a vanishing divergence $h_{ij}\phi_{,ij}\phi=(h_{ij}\phi_{,i})_{,j}-h_{ij,j}\phi_{,i}\phi-h_{ij}\phi_{,i}\phi_{,j}=-h_{ij}\phi_{,i}\phi_{,j}$, so the scaler-tensor interaction coming from gravitational nonlinearity is
\begin{equation}
\mathcal{L}_{S-T \ int}=\frac{1}{8\pi G}a^2h_{ij}\phi_{,i}\phi_{,j}=M^2_pa^2h_{ij}\phi_{,i}\phi_{,j}
\end{equation}
\subsection{Interaction with scaler inflation field perturbation}
The tensor environment also interacts with the perturbed inflation field, which induces a back reaction on the metric. In our paper, we study a single scaler field inflation minimal coupled with gravity
\[\mathcal{L}=\mathcal{L}_{gra}+\mathcal{L}_{inflation}\]
where
\[\mathcal{L}_\varphi=-N\sqrt{g}\left[\frac{1}{2}\partial_\mu\varphi g^{\mu\nu}\partial_\nu\varphi+V(\varphi)\right]\]
Insert the perturbation field $\varphi=\varphi_0+\delta\varphi$ and expand to third order
\begin{equation}
\mathcal{L}_{\varphi-T \ int}=\frac{1}{2}a^2h_{ij}\delta\varphi_{,i}\delta\varphi_{,j}
\end{equation}

\section{Calculation of trace tensor}\label{apdx:calculate trace tensor}
Our task is to calculate partial trace $Tr_T(T_{\bm{x}}T_{\bm{x}'}\varrho_T)$. We use the occupation number basis $\{\bigotimes_{\bm{k},s=+,\times}|n(\bm{k},s)\rangle\}$ of the initial BD vacuum state to expand the partial trace. Because the time evolution is unitary, this basis is still orthonormal in later time though it's no longer eigenstates of the Hamiltonian. Under such basis
\[
\begin{aligned}
Tr_T(T_{\bm{x}}T_{\bm{x}'}\varrho_T)&=\sum_{m(\bm{p},s),n(\bm{q},r)}\langle m(\bm{p},s)|T_{\bm{x}}(t)T_{\bm{x}'}(t')|n(\bm{q},r)\rangle p[n]\langle n(\bm{q},r)|m(\bm{p},s)\rangle\\&=\sum_{n(\bm{q},r)}p[n]\langle n(\bm{q},r)|T_{\bm{x}}(t)T_{\bm{x}'}(t')|n(\bm{q},r)\rangle\\&=\sum_{n}p[n]N[n]\langle0|\left[\prod_{\bm{p},s}(a_{\bm{p}}^s)^{n(\bm{p},s)}\right]T_{\bm{x}}T_{\bm{x}'}\left[\prod_{\bm{q},r}(a_{\bm{q}}^{r\dagger})^{n(\bm{q},r)}\right]|0\rangle
\end{aligned}
\]
where $n(\bm{k},s)$ is the occupation number in the space labeled by $(\bm{k},s)$ and $\varrho_T=p[n]|n\rangle\langle n|$. $N[n]$ is the normalization factor. Insert expansion \eqref{T_x,S_x} one get
\[\begin{aligned}
\sum_{n}p[n]\langle n|T_{\bm{x}}T_{\bm{x}'}|n\rangle=&\sum_np[n]N[n]\langle0|\left[\prod_{\bm{p},r}(a_{\bm{p}}^r)^{n(\bm{p},r)}\right]\int\frac{dk^3}{(2\pi)^3}\frac{dk'^3}{(2\pi)^3}\sum_{s,s'}\epsilon_{ij}^s(\bm{k})\epsilon_{kl}^{s'}(\bm{k}')e^{i(\bm{k}\cdot\bm{x}+\bm{k}'\cdot\bm{x}')}\\&\times(u_k(t)a^s_{\bm{k}}+u_k^*(t)a_{-\bm{k}}^{s\dagger})(u_{k'}(t')a^{s'}_{\bm{k}'}+u_{k'}^*(t')a_{-\bm{k}'}^{s'\dagger})\left[\prod_{\bm{p}',r'}(a_{\bm{p}'}^{r'\dagger})^{n(\bm{p}',r')}\right]|0\rangle\\=&\sum_np[n]N[n]\langle0|\left[\prod_{\bm{p},r}(a_{\bm{p}}^r)^{n(\bm{p},r)}\right]\int\frac{dk^3}{(2\pi)^3}\frac{dk'^3}{(2\pi)^3}\sum_{s,s'}\epsilon_{ij}^s(\bm{k})\epsilon_{kl}^{s'}(\bm{k}')e^{i(\bm{k}\cdot\bm{x}+\bm{k}'\cdot\bm{x}')}\\&\times(u_ku^*_{k'}a^s_{\bm{k}}a^{s'\dagger}_{-\bm{k}'}+u_k^*u_{k'}a^{s\dagger}_{-\bm{k}}a^{s'}_{\bm{k}'})\left[\prod_{\bm{p}',r'}(a_{\bm{p'}}^{r'\dagger})^{n(\bm{p}',r')}\right]|0\rangle\\=&\sum_np[n]\int\frac{dk^3}{(2\pi)^3}\frac{dk'^3}{(2\pi)^3}\sum_{s,s'}\epsilon_{ij}^s(\bm{k})\epsilon_{kl}^{s'}(\bm{k}')\left[u_ku^*_{k'}\frac{N[n]}{\sqrt{N[n_1]N[n_1']}}\langle n_1|n_1'\rangle\right.\\&\left.+u_k^*u_{k'}\frac{N[n]}{\sqrt{N[n_2]N[n_2']}}\langle n_2|n_2'\rangle-u_k^*u_{k'}(2\pi)^3\delta^{ss'}\delta^{(3)}(\bm{k}+\bm{k}')\right]e^{i(\bm{k}\cdot\bm{x}+\bm{k}'\cdot\bm{x}')}
\end{aligned}\]
where function $n_1,n'_1$ are defined as
\[n_1(\bm{p},r)=\left\{\begin{aligned}
&n(\bm{k},s)+1,\ \bm{p}=\bm{k},r=s\\
&n(\bm{p},r),\ \ otherwise
\end{aligned}\right.\qquad n_1'(\bm{p},r)=\left\{\begin{aligned}
&n(-\bm{k'},s')+1,\ \bm{p}=-\bm{k'},r=s'\\
&n(\bm{p},r),\ \ otherwise
\end{aligned}\right.\]
$n_2,n'_2$ are similarly defined. Normalization factor $N[n]=\prod_{\bm{k},s}n(\bm{k},s)!$ is infinite but quotient $N[n]/N[n_1]$ is finite
\[\frac{N[n]}{\sqrt{N[n_1]N[n_1']}}=\frac{N[n]}{N[n_1]}=n(\bm{k},s)+1\]
Further utilize the orthogonality of basis $\{|n\rangle\}$ one get
\[
\begin{aligned}
Tr_T[T_{\bm{x}}T_{\bm{x}'}\varrho_T]=&\sum_np[n]\int\frac{dk^3}{(2\pi)^3}\sum_s\epsilon_{ij}^s(\bm{k})\epsilon_{kl}^s(\bm{k})e^{i\bm{k}\cdot(\bm{x}-\cdot\bm{x}')}\\&\times[(n(\bm{k},s)+1)u_k(t)u_k^*(t')+n(\bm{k},s)u_k^*(t)u_k(t')]
\end{aligned}
\]
Specialize to the BD vacuum case (i.e: ground states in every $(\bm{k},s)$ subspace $p[0]=1$)
\begin{equation}\label{Trace vacuum}
Tr_T(T_{\bm{x}}T_{\bm{x}'}\varrho_T)=\int\frac{dk^3}{(2\pi)^3}\sum_s\epsilon_{ij}^s(\bm{k})\epsilon_{kl}^s(\bm{k})u_k(t)u_{k}^*(t')e^{i\bm{k}\cdot(\bm{x}-\bm{x}')}
\end{equation}
Change to spherical coordinates
\[\begin{aligned}
Tr_T(T_{\bm{x}}T_{\bm{x}'}\varrho_T)=\frac{1}{(2\pi)^3}\int_0^{+\infty} k^2dku_k(t)u_{k}^*(t')\int \sin\theta d\theta d\varphi e^{ik|\bm{x}-\bm{x}'|\cos\theta}\sum_s\epsilon_{ij}^s(\theta,\varphi)\epsilon_{kl}^s(\theta,\varphi)
\end{aligned}\]
Let's start from the angular part. Polarization basis for a wave propagates in the $\hat{z}$ direction
\[\epsilon^+=\frac{1}{\sqrt{2}}\begin{pmatrix}
1&0&0\\0&-1&0\\0&0&0
\end{pmatrix}\qquad
\epsilon^\times=\frac{1}{\sqrt{2}}\begin{pmatrix}
0&1&0\\1&0&0\\0&0&0
\end{pmatrix}\]
It can be rotated to an arbitrary direction $\hat{\bm{k}}=(\sin\theta\cos\varphi,\sin\theta\sin\varphi,\cos\theta)$ by
\[\epsilon^s(\theta,\varphi)=R(\theta,\hat{\bm{n}})^T\epsilon^sR(\theta,\hat{\bm{n}})\]
where $\hat{\bm{n}}=(-\sin\varphi,\cos\varphi,0)$. There are other possible rotations satisfying $\hat{\bm{z}}\cdot R=\hat{\bm{k}}$ but they are all equivalent since the polarization tensors are rotational invariant in the plane orthogonal to $\hat{\bm{k}}$. Thus the angular integral yields
\[\begin{bmatrix}
    \begin{pmatrix}a&0&0\\0&b&0\\0&0&e\end{pmatrix}
    &\begin{pmatrix}0&d&0\\d&0&0\\0&0&0\end{pmatrix}
    &\begin{pmatrix}0&0&c\\0&0&0\\c&0&0\end{pmatrix}\\
    \begin{pmatrix}0&d&0\\d&0&0\\0&0&0\end{pmatrix}
    &\begin{pmatrix}b&0&0\\0&a&0\\0&0&e\end{pmatrix}
    &\begin{pmatrix}0&0&0\\0&0&c\\0&c&0\end{pmatrix}\\
    \begin{pmatrix}0&0&d\\0&0&0\\c&0&0\end{pmatrix}
    &\begin{pmatrix}0&0&0\\0&0&c\\0&c&0\end{pmatrix}
    &\begin{pmatrix}e&0&0\\0&e&0\\0&0&-2e\end{pmatrix}
\end{bmatrix}\]
where the $\hat{z}$ direction si chosen to be $\Delta\bm{x}=\bm{x}-\bm{x}'$ and
\[\begin{aligned}
&a=2\pi\left[\frac{r(2r^2-9)\cos r+(r^4-5r^2+9)\sin r}{r^5}\right]\\
&b=-2\pi\left[\frac{r(2r^2+3)\cos r+(r^4-r^2-3)\sin r}{r^5}\right]\\
&c=-4\pi\left[\frac{r(r^2-6)\cos r+(-3r^2+6)\sin r}{r^5}\right]\\
&d=(a-b)/2 \qquad e=-a-b
\end{aligned}\]
\[r=k|\bm{x}-\bm{x}'|=k\Delta x\]
Notice that $a\to 16\pi/15, \ b\to-8\pi/15, \ c\to4\pi/5$ when $r\to0$, then the radial integration diverges at $k=0$ since mode functions $u_k(t)\propto H^{(1)}_{\nu}(kt)\sim k^{-\nu}$. To avoid divergence near zero, we assume that the tensor environment contains only well-defined particles (i.e: $kt\gtrsim1$). We will further treat the environment roughly as plane waves since argument of Hankel $kt\gtrsim1$
\[k^2u_k(t)u^*_k(t')\to \frac{k}{2}e^{-ik(t-t')}\]
We can extend the integration limit back to $0<k<+\infty$ after approximating mode functions by plane waves. The radial integrals involving $a$ and $b$ still posses some singularity (i.e: $\sin r/r$)
\[\frac{\pi}{\Delta x}\int_0^{+\infty}dke^{-ik(t-t')}\sin(k\Delta x)\]
where $+$ for $a$ and $-$ for b. Then use Sokhotski–Plemelj theorem
\[\begin{aligned}
    &\int_0^{+\infty}dk\frac{\pi}{\Delta x}e^{-ik(t-t')}\sin (k\Delta x)=\lim_{\varepsilon\to0^+}\int_0^{+\infty}dk\frac{\pi}{\Delta x}e^{-ik(t-t')-k\varepsilon}\sin (k\Delta x)\\&=-\frac{\pi}{2\Delta x}\lim_{\varepsilon\to0^+}\left(\frac{1}{t-t'-\Delta x-i\varepsilon}-\frac{1}{t-t'+\Delta x-i\varepsilon}\right)\\
    &=-\frac{i\pi^2}{2\Delta x}\delta(t-t'-\Delta x)-\pi PV\left(\frac{1}{(t-t')^2-\Delta x^2}\right)
\end{aligned}\]
The rest of the integration, singularity extracted, yields analytic expression when $t-t'>\Delta x$
\[\begin{aligned}
&a\to \frac{\pi}{\Delta x^2}\left[-1-\frac{3}{2}s^2-\frac{1}{4}s(1+3s^2)\log\left(\frac{s-1}{s+1}\right)\right]\\
&b\to-\frac{\pi}{\Delta x^2}\left[-\frac{7}{3}+\frac{1}{2}s^2+\frac{1}{4}s(-5+s^2)\log\left(\frac{s-1}{s+1}\right)\right]\\
&c\to-\frac{2\pi}{\Delta x^2}\left[-\frac{1}{3}-s^2-\frac{1}{2}s^3\log\left(\frac{s-1}{s+1}\right)\right]
\end{aligned}\]
where $s=\Delta t/\Delta x$. These expressions hold within the light-cone $s>1$. Some common properties
\begin{enumerate}
	\item They diverge at $t-t'-\Delta x=0$ as $\log(t-t'-\Delta x)$;
	\item They decays like $(t-t'-\Delta x)^{-2}$ near infinity;
	\item Their integration $\int^{\infty}_1 ds$ converge.
\end{enumerate}
Notice that
\[5a+b+4c=8\pi\frac{\sin r}{r}\]  We will see later that it is the key property that makes these nonsingular explicit expressions do not contribute to decoherence in the leading order. However, we keep track of them for now. Let's propose the $\delta$ function approximation
\[\begin{aligned}
&a\to\alpha_1\frac{3\pi}{4\Delta x}\delta(t-t'-\Delta x)\\
&b\to-\alpha_2\frac{13\pi}{12\Delta x}\delta(t-t'-\Delta x)\\
&c\to-\alpha_4\frac{2\pi}{3\Delta x}\delta(t-t'-\Delta x)
\end{aligned}
\]
where factor $\alpha_i(kt,\Delta t)=|\alpha_i|e^{i\delta_i}$. Fig-2 shows the norm and phase of $\alpha$ near horizon exit. 
\begin{figure}\label{alpha}
    \centering
    \subfigure[Norm of $\alpha$]{\includegraphics[width=3in]{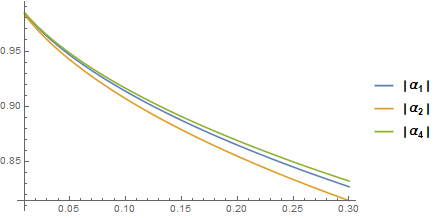}}
    \subfigure[Phase of $\alpha$]{\includegraphics[width=3in]{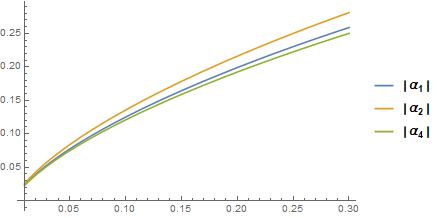}}
    \caption{Graph of $\alpha$ near $kt\sim 1$, calculated by using $u_k(t)u^*_k(t)$ as the integration kernel. The x-axis is $\Delta x$.}
\end{figure}

We may further absorb the principle value term into the $\delta$ function approximation by 
\[PV\left(\frac{1}{(t-t')^2-\Delta x^2}\right)=-\frac{p}{\Delta x}\delta(t-t'-\Delta x)\]
Fig-3 illustrates numerical plot of $p\sim-\Delta x PV\left(\frac{1}{(t-t')^2-\Delta x^2}\right)(u_k(t)u^*_k(t))/(u_k(t-\Delta x)u^*_k(t-\Delta x))$ near $|kt|=1$.
\begin{figure}\label{p}
	\centering
	\includegraphics[width=3.5in]{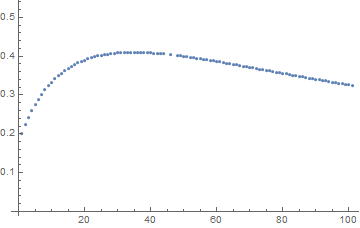}
	\caption{Plot of $p$ at $t=-5, \ k=0.2$ with x-coordinate $x=10\Delta x$.}
\end{figure}
Collect all the results and insert into \eqref{Trace vacuum} one obtain $ Tr_T(T_{\bm{x}}(t)T_{\bm{x}'}(t')\varrho_T)\equiv T_{ijkl}\delta(t-t'-\Delta x)/\Delta x$ where $T_{ijkl}$ is given by \eqref{T trace tensor}.

\section{Field amplitude basis}\label{apdx:field amplitude basis}
Our goal is to construct a state $|\bm{\xi}\rangle$ from the vacuum state $|0\rangle$ such that
\[\hat{\phi}(\bm{x})|\bm{\xi}\rangle=\xi(\bm{x})|\bm{\xi}\rangle\]
where $\hat{\phi}(\bm{x})$ is the field operator of some arbitrary scaler field and $\xi(\bm{x})$ is the classical field amplitude at $\bm{x}$.

We start from a simple harmonic oscillator case. Let $|0\rangle$ be the ground state then define
\[|0_x\rangle=\exp(-a^\dagger a^\dagger/2)|0\rangle\]
where $a^\dagger$ is the creation operator. One can easily verify using $[a,a^\dagger]=1$ that $|0_x\rangle$ is an eigenstate of position operator $\hat{x}|0_x\rangle=0$. Thus an any position eigenstates may be generated by displacement
\[|x\rangle=e^{-i\hat{p}x}e^{-a^\dagger a^\dagger/2}|0\rangle\]
Extend to quantum field
\[|\bm{\xi}\rangle=\prod_{\bm{x}}\exp(-i\hat{\pi}(\bm{x})\xi(\bm{x}))|0_\phi\rangle\]
and
\[|0_\phi\rangle=\prod_{\bm{x}}\exp(-\phi^+(\bm{x})\phi^+(\bm{x})/2)|0\rangle\]
where $\phi^+(\bm{x})=[\phi^-(\bm{x})]^\dagger$ is the creation operator. The annihilation operators are defined as usual
\[\phi^-(\bm{x})\equiv\frac{\hat{\phi}(\bm{x})+i\hat{\pi}(\bm{x})}{\sqrt{2}}\]
Momentum operator $\hat{\pi}$ acts on a wave functional as
\[\langle\bm{\xi}|\hat{\pi}(\bm{x})|\bm{\Psi}\rangle=-iM^3_p\frac{\delta}{\delta\xi(\bm{x})}\langle\bm{\xi}|\bm{\Psi}\rangle\]
where $M_p^3$ comes from the fact that $\hat{\pi}$ is momentum density.

Specialize to our scaler field $\hat{v}$, we need to know how the creation and annihilation operators $b_{\bm{k}},b^\dagger_{\bm{k}}$ acts on the field amplitude basis. First define Fourier transform $\mathcal{F}[\bm{v}](\bm{k})=\int d\bm{x}\hat{v}(\bm{x})e^{-i\bm{k}\cdot\bm{x}}$ and use \eqref{mode expansion of scaler field}
\begin{displaymath}
\begin{bmatrix}
\mathcal{F}[\hat{\bm{v}}](\bm{k})\\\mathcal{F}[\hat{\bm{\pi}}](\bm{k})
\end{bmatrix}=\begin{bmatrix}
v_k&v^*_k\\v'_k&v^{\prime*}_k
\end{bmatrix}\begin{bmatrix}
\hat{b}_{\bm{k}}\\\hat{b}^\dagger_{-\bm{k}}
\end{bmatrix}
\end{displaymath}
The solution is
\begin{displaymath}
\begin{bmatrix}
\hat{b}_{\bm{k}}\\\hat{b}^\dagger_{-\bm{k}}
\end{bmatrix}=\frac{1}{\mathcal{W}(v_k,v^*_k)}\begin{bmatrix}
v^{\prime*}_k&-v^*_k\\-v'_k&v_k
\end{bmatrix}\begin{bmatrix}
\mathcal{F}[\hat{\bm{v}}](\bm{k})\\\mathcal{F}[\hat{\bm{\pi}}](\bm{k})
\end{bmatrix}=-i\begin{bmatrix}
v^{\prime*}_k&-v^*_k\\-v'_k&v_k
\end{bmatrix}\begin{bmatrix}
\mathcal{F}[\hat{\bm{v}}](\bm{k})\\\mathcal{F}[\hat{\bm{\pi}}](\bm{k})
\end{bmatrix}
\end{displaymath}
where we have used the normalization condition \eqref{mode normalization} in the last equality. Operator $\mathcal{F}[\hat{\bm{v}}](\bm{k}),\mathcal{F}[\hat{\bm{\pi}}](\bm{k})$ acting on field amplitude state yields
\[\mathcal{F}[\hat{\bm{v}}](\bm{k})|\bm{\xi}\rangle=\mathcal{F}[\xi](\bm{k})|\bm{\xi}\rangle\qquad \langle\bm{\xi}|\mathcal{F}[\hat{\bm{\pi}}](\bm{k})|\bm{\Psi}\rangle=-i\mathcal{F}[\frac{\delta\Psi[\xi]}{\delta\xi}](\bm{k})\]

\section{Integration involving $T_{ijkl}$}\label{apdx:integration involving T}
In this section we calculate several integrations involving $T_{ijkl}$ and prove that $\Gamma$ is positive definite in the meantime. It suffices to calculate integral of the form
\[\int d\hat{\bm{s}}'T_{ijkl}(\hat{\bm{s}}')u_iu_jv_kw_l\]
where $\bm{u},\bm{v},\bm{w}$ are three arbitrary vectors. Since it's a spherical integration and $T_{ijkl}(\hat{\bm{r}}')$ is a spacial tensor, the final result must be constructed by scaler contraction of these vectors: $\bm{u}\cdot\bm{v}$ and $|\bm{u}\times\bm{v}|$. The module of cross product is possible in that $|\bm{u}\times\bm{v}|=|u||v|\sin\langle\bm{u},\bm{v}\rangle$ only depends on the relative direction of $\bm{u}$ and $\bm{v}$. For simplicity, we can choose a frame with $\bm{u}=u(0,0,1)$ being the z-direction and $\bm{v}=v(\sin\theta_{uv},0,\cos\theta_{uv})$ residing in the xz-plane. Vector $\bm{w}=w(\sin\theta_{uw}\cos\varphi_{uw},\sin\theta_{uw}\sin\varphi_{uw},\cos\theta_{uw})$. The result is
\[\frac{4\pi}{15}(5T_1+T_2+4T_4)u^2vw(2\cos\theta_{uv}\cos\theta_{uw}-\cos\varphi_{uw}\sin\theta_{uv}\sin\theta_{uw})\]
Cast it into a scaler contraction form
\[\int d\hat{\bm{s}}'T_{ijkl}(\hat{\bm{s}}')u_iu_jv_kw_l=\frac{4\pi}{15}(5T_1+T_2+4T_4)[3(\bm{u}\cdot\bm{v})(\bm{u}\cdot\bm{w})-(\bm{u\cdot\bm{u}})(\bm{v}\cdot\bm{w})]\]
Recall the leading term in decoherence, let $\bm{u}=\nabla\xi$ and $\bm{v}=\bm{w}=\nabla\zeta$, omitting bars for dimensionless variables. The real part then writes
\[\frac{8\pi}{15}\left[\left(|\nabla\xi|^2-|\nabla\zeta|^2\right)^2+3|\nabla\xi\times\nabla\zeta|^2\right]\int_{\Delta} ds' s'\Re(5T_1+T_2+4T_4)\]
Notice we have commented before $5a+b+4c=8\pi\sin (k\Delta x)/(k\Delta x)$ so the $\Delta$ integration is $p\Delta^2/(4\pi^2)$. Thanks to the positive principle value $p$, we indeed have a positive definite decoherence rate.

The linear correction term in \eqref{The Master Eq dimensionless} is easily calculated by letting $\bm{u}=\nabla\xi, \bm{v}=\nabla \zeta,\bm{w}=\hat{\bm{s}}$.
Finally in \eqref{The Master Eq matrix simplified} the term $\partial^{(-)}_{ij}[\xi,\zeta](\bm{r}+\bm{r}')$ contains an first order integral
\[\int d\hat{\bm{s}}'T_{ijkl}(\hat{\bm{s}}')\hat{\bm{s}}'\cdot\nabla(\partial_i\xi\partial_j\xi)\partial_k\zeta\partial_l\zeta\]
This term vanishes. The reason is that the result should be a scaler constructed by dot contraction, but what we have here are $H_{\bar{\xi}}$ the Hessian matrix, $\nabla\xi$ and two $\nabla\zeta$. Observe that it's impossible to contract one matrix and three vectors to give a scaler, thus the integral must vanish.

\end{document}